\RequirePackage{ifpdf}
\ifpdf 
\documentclass[pdftex]{sigma}
\else
\documentclass{sigma}
\fi

\renewcommand{\arraystretch}{1.7}
\newtheorem{prop}[theorem]{Proposition}

\newenvironment{pf}{\begin{proof} \rm}{\hfill$\square$ \end{proof}}
\newcounter{appendix}
\newcommand{\CH}{{\cal H}}

\newcommand{\CS}{{\cal S}}

\newcommand{\CE}{{\cal E}}

\newcommand{\CM}{{ M}}

\newcommand{\CP}{{\cal P}}

\makeatletter \@addtoreset{equation}{section}

\makeatother

\newcommand{\Jj}{{\mathsf J}}
\newcommand{\Nn}{{\mathsf N}}

\newcommand{\Ee}{{\mathsf E}}

\newcommand{\uu}{{\mathsf u}}

\newcommand{\Pp}{{\mathsf P}}

\newcommand{\Qq}{{\mathsf Q}}

\newcommand{\RR}{{\mathbb R}}

\def\ga{\gamma}         
\def\be{\beta}
\def\al{\alpha}

\def\la{\lambda}

\def\dsl{\displaystyle}

\newcommand{\del}{{\partial}}

\def\parpo#1#2{\{#1,#2\}}

\def\mat2#1#2#3#4{{\left(\begin{array}{cc}#1 & #2\\ #3 & #4
      \end{array}\right)}}

\def\mats2#1#2#3#4{{\left(\begin{array}{cc}#1 & #2\vspace{2truemm} \\ #3 & #4
\end{array}\right)}}

\def\ddd#1#2{\displaystyle{\frac{\partial #1}{\partial #2}}}

\def\Nij{Nijenhuis}

\def\alg{{\mathfrak g}}

\def\dncoo{Darboux--Nijenhuis co-ordinates}

\def\St{St\"ackel}

\newcommand{\Lie}[1]{{\text{Lie}_{{#1}}}}
\newcommand{\so}{{\mathfrak{so}}}

\begin{document}

\allowdisplaybreaks

\renewcommand{\PaperNumber}{032}

\FirstPageHeading

\renewcommand{\thefootnote}{$\star$}

\ShortArticleName{A Note on the Rotationally Symmetric $SO(4)$
Euler Rigid Body}

\ArticleName{A Note on the Rotationally Symmetric\\
$\boldsymbol{SO(4)}$ Euler Rigid Body\footnote{This paper is a
contribution to the Vadim Kuznetsov Memorial Issue `Integrable
Systems and Related Topics'. The full collection is available at
\href{http://www.emis.de/journals/SIGMA/kuznetsov.html}{http://www.emis.de/journals/SIGMA/kuznetsov.html}}}

\Author{Gregorio FALQUI}

\AuthorNameForHeading{G. Falqui}

\Address{Dipartimento di Matematica e Applicazioni,
Universit\`a di Milano -- Bicocca,\\
via R. Cozzi, 53, 20125 Milano, Italy}
\Email{\href{mailto:gregorio.falqui@unimib.it}{gregorio.falqui@unimib.it}}
\URLaddress{\url{http://www.matapp.unimib.it/~falqui}}

\ArticleDates{Received November 15, 2006, in f\/inal form February
02, 2007; Published online February 26, 2007}

\Abstract{We consider an $SO(4)$ Euler rigid body with two
`inertia momenta' coinciding. We study it from the point of view
of bihamiltonian geometry. We show how to algebraically integrate
it by means of the method of separation of variables.}


\Classification{37K10; 70H20; 14H70}

\renewcommand{\thefootnote}{\arabic{footnote}}
\setcounter{footnote}{0}

\section{Introduction} It is fair to say that the problem of the free rigid
body in $\RR^4$ (i.e., the $SO(4)$ Euler or Euler--Manakov rigid
body or top\footnote{Actually, this model was studied, in the XIX
century, by Schottky and K\"otter.}) can be still considered as an
interesting problem in the theory of Separation of Variables (SoV)
for Hamilton--Jacobi equations.

Ingenious methods were devised to solve the very classical problem
of the Euler--Manakov top, (see, e.g., \cite{AvM83,Audin, BBEIM,
Bo86,Ma76,MiFo78,Ra82, VeNo84}, but this list -- as well as the
bibliography of the present paper -- is by far incomplete), in
particular to characterise the solutions by means of suitable
techniques coming from algebraic geometry.

In this paper we will consider, from the point of view of
bihamiltonian geometry, a degenerate case of Euler top that can be
called rotationally symmetric. Namely, we consider the case in
which two of the four independent parameters that enter the
construction (that, after Manakov's construction, can be termed
generalised principal inertia moments) coincide. This is a
particular case of the general $SO(4)$ system, but it possesses
some interesting and peculiar features that deserve, in the
author's opinion, to be spelled out.

It should be noticed that in \cite{SkTa99} the SoV problem was
implicitly considered in the study of SoV for elliptic Gaudin
models, and basically solved, with a considerable amount of hard
computations and ingenuity. Furthermore, in \cite{MaSo05},
interesting classes of solutions to the Hamilton--Jacobi equations
associated with the $SO(4)$ Euler--Manakov systems (as well as
other alike systems) were recently obtained. In the present paper
we will follow a more direct and geometrical way to approach it,
with the aim of giving a simple and explicit setting for the
Separation of Variables problem of this symmetric SO(4) Euler top.

The framework we will use to study this system is the so-called
bihamiltonian setting for Separation of Variables for
Gel'fand--Zakharevich \cite{GZ00} systems that has been introduced
a few years ago (see, e.g., \cite{Bl99,FMT98,MT97}, and references
quoted therein) and formalised in \cite{FaPe03}; we will sketch
the content of this method in Section \ref{sec.1}.

After that, in Section \ref{sec.2} we will brief\/ly resume those
aspects of the $SO(4)$ Euler--Manakov system that are relevant in
our analysis, and in the core of the paper (Section \ref{sec.3})
we will consider the rotationally symmetric case and solve,
applying the recipes described in Section \ref{sec.1}, the SoV
problem for this Hamiltonian system.

Separation of Variables, both in classical and quantum systems,
was a research arena in which Vadim Kuznetsov obtained notable
results. In particular, his papers
\cite{Ku92,KuNiSk97,Ku02,KuPeRa04} were deeply inf\/luential on
the study of the SoV problem (and more general integrability) from
the point of view of bihamiltonian geometry, carried out by the
author of the present paper in collaboration with F. Magri, M.
Pedroni, and G. Tondo. We had the opportunity to discuss with him
these and related subjects many times, and appreciate his
scientif\/ic, as well as human talents. With deep sorrow I
dedicate this work to his memory.

\section{A bihamiltonian set-up for SoV}\label{sec.1}
In this Section we will brief\/ly review a scheme for solving the
SoV problem in the Hamilton Jacobi equations, based on properties
of bihamiltonian manifolds. We will discuss those features that
are relevant for the case at hand, referring to \cite{FaPe03} and
to \cite{Fa03} for a comprehensive theoretical presentation, as
well as for a wider list of references.

Let $(\CM, P_1,P_2)$ be a bihamiltonian manifold, that is, a
manifold endowed with a pair of compatible Poisson brackets
$\{\cdot,\cdot\}_{P_i}$, $i=1,2$, or, equivalently, with two
compatible Poisson bivectors (or operators) $P_1$, $P_2$, related
to the brackets by the well-known formulas
\begin{gather*}
  \{f,g\}_{P_i}= \langle df,P_i dg\rangle \qquad \forall\,f,g\in
  C^\infty(\CM), \quad i=1,2.
\end{gather*}
We consider a Gel'fand--Zakharevich bihamiltonian system with one
bihamiltonian chain. That is, we consider the datum, on the
bihamiltonian manifold $(\CM, P_1,P_2)$, of an `anchored'
Lenard--Magri chain of length $n>0$,
\begin{gather}
\label{eq:1.1} P_1 dH_0=0, \qquad P_1 dH_i=P_2 d H_{i-1},\quad
i=1,\ldots, n, \qquad
 P_2 dH_{n}=0,
\end{gather}
that is, a family of bihamiltonian vector f\/ields originating
from a Casimir function of one Poisson operator and ending in a
Casimir function of the other Poisson operator. We may suppose
that $p$ additional Casimir functions common to the two structures
$C_1,\ldots,C_p$ be also present\footnote{In the $SO(4)$
  Euler top, indeed, we will have one of such Casimir functions.},
and we require that the system be complete, i.e., that
$n=\mathop{\rm dim}M-1-p$ independent vector f\/ields f\/ill in
the chain \eqref{eq:1.1}.

Such a system provides (families of) Liouville integrable systems
as follows. One considers a (generic) symplectic leaf $\CS$ of one
of the two Poisson operators, say, $P_1$.

$\CS$ is  a submanifold of $\CM$ def\/ined by f\/ixing the values
of all Casimir functions $H_0, C_1,\ldots, C_p$ of~$P_1$. Any of
the vector f\/ields of the chain \eqref{eq:1.1}, say $X_1=P_1 dH_1
(=P_2 dH_0))$, restricts to $\CS$, is still Hamiltonian with
respect to (the restriction of) the chosen Poisson operator $P_1$
(which becomes an ordinary symplectic operator on $\CS$), and,
thanks to the basic property of Lenard Magri chains, comes
equipped with the right number of involutive integrals, namely the
(restriction to $\CS$ of) the other Hamiltonian functions of the
vector f\/ields $X_i$, $i=2,\ldots, N$ of~\eqref{eq:1.1}.

What is lost, in general, in this procedure, is the bihamiltonian
structure of the equations: indeed, the second Poisson operator
does not restrict to $\CS$, since it does not leave the
function~$H_0$, which is a Casimir of $P_0$ invariant, i.e.,
Hamitonian vector f\/ields generated by $P_2$ do not leave the
submanifolds $\CS$ invariant.

However, as it is shown
in \cite{FaPe03}, the manifold $M$ may be provided with another suitable bihamiltonian
structure. That is, along with $P_1$,  a ``new'' second Poisson operator $Q$
can be def\/ined deforming $P_2$;  $Q$
can be restricted to the symplectic leaves $\cal{S}$ and has
those properties needed for integrating the vector fields of the chain (\ref{eq:1.1}).

For this new structure to exist,  conditions are to be
fulf\/illed. Namely, one has to f\/ind a~vector f\/ield $Z$,
def\/ined on $M$, such that:

i)~$Z$ is a  symmetry of $P_1$, transversal to the submanifolds
$H_0={\rm const}$,
  and leaving the common Casimir functions $C_i$ invariant:
\begin{gather*}
\Lie{Z}(P_1)=0,\qquad \Lie{Z}(H_0)=1,\qquad \Lie{Z}(C_i)=0,\quad
i=1,\ldots, p.
\end{gather*}

ii)~It deforms of the second Poisson tensor $P_1$ as follows:
\begin{gather*}
\Lie{Z}(P_2)=Y\wedge Z, \quad \text{for some vector f\/ield} \ Y.
\end{gather*}

Indeed, under these conditions it holds that the bivector
$Q:=P_2-X_1\wedge Z$ (where $X_1$ is the f\/irst vector f\/ield of
the Lenard--Magri chain \eqref{eq:1.1}) satisf\/ies:

1.~$Q$ is a Poisson structure on $M$, compatible with $P_1$ which
shares with $P_1$ all the Casimirs, and hence, together with $P_1$
induces a bihamiltonian structure on the symplectic leaves $\CS$.

2.~The Hamiltonians $H_i$ do not form anymore Lenard--Magri
sequences
  w.r.t.\ this new Poisson pair $(P_1,Q)$, but still are in involution also
  w.r.t.\ the deformed (or new) structure $Q= P_2-X_1\wedge Z$.

These two properties are very important for our purposes; indeed,
from the f\/irst one it follows \cite{GZ93,Ma90} that the pair
$(P_1, Q)$ def\/ines, on each symplectic leaf $\CS$, a special set
of co-ordinates, called {\em
  Darboux--Nijenhuis} co-ordinates, associated with the eigenvalues of a
torsionless `recursion' operator.

From the second one, it follows \cite{FaPe03} that the
Hamiltonians $H_i$ are separable in these DN co-ordinates, that
is, that the Hamilton--Jacobi equation associated with any of the
Hamiltonians~$H_i$ is separable in these co-ordinates. Hence, the
f\/irst and basic step of the bihamiltonian recipe for SoV of
Gel'fand--Zakharevich systems of type \eqref{eq:1.1} essentially
boils down to f\/ind/guess this vector f\/ield $Z$, which will be
referred to as the {\em transversal} vector f\/ield.

For the reader's convenience, as well as to provide the necessary
background to the calculations presented in Section \ref{sec.3},
we discuss more in details the construction of DN co-ordinates. A
preliminary remark is in order. As the two structures $P_1$ and
$Q$ share the same symplectic leaves, we will generically use the
same letters $P_1$, $Q$ also for their natural restrictions to the
symplectic leaves to avoid cumbersome notations.

As already noticed, on any (generic) symplectic leaf of $P_1$ this
operator is symplectic and thus invertible; hence, the
compositions
\begin{gather*}
  N:=QP^{-1}\colon T\CS\to T\CS, \qquad\text{and}\qquad
N^*:=P_1^{-1}Q\colon T^*\CS\to T^*\CS
\end{gather*}
are well def\/ined. In the literature, $N$ is called a \Nij\ or
recursion or hereditary operator; in our setting, its adjoint
operator, $N^*$ will play a more visible role.

Being the ratio of two antisymmetric operators, $N^*$ has up to
$m=(1/2)\mathop{\rm dim} \CS$ distinct eigenvalues
$\la_1,\ldots,\la_m$. Under the  assumption that the number of
these distinct eigenvalues be exactly $m$, it follows, basically
as a consequence of the compatibility condition between $P_1$
and~$Q$, that, for each $\la_i$ there is a pair of canonical
(w.r.t.~$P_1$) co-ordinates $f_i$, $g_i$ that generates the
eigenspaces of $N^*$, that is, it holds
\begin{gather}
\label{eq:1.6} N^* df_i=\la_i\, df_i;\qquad N^* dg_i, \qquad
\{f_i,g_j\}_{P_1}=\delta_{ij}.
\end{gather}
These co-ordinates are called Darboux--Nijenhuis co-ordinates
associated with the pair~$P_1$,~$Q$. Notice that the `eigenvalue'
relations written above imply the Poisson bracket relations $
\{f_i,g_j\}_Q$ $=\delta_{ij}\la_i $ w.r.t.\ the second Poisson
structure. Fortunately enough, in generic cases, there is no need
to integrate (all the) two-dimensional distributions $\mathop{\rm
Ker}(N^*-\la_i)$ to actually f\/ind DN co-ordinates, thanks to the
following results.

It should be noticed that, since $N^*$ depends on the point of
$\CS$, its eigenvalues are functions def\/ined on $\CS$. In
particular, non-constant eigenvalues of $N^*$ do provide DN
co-ordinates. This means that, if $d\la_i\neq 0$, then one can
choose the function $\la_i$ as the co-ordinate function $f_i$
of~\eqref{eq:1.6}.

To f\/ind the missing canonical co-ordinate associated with the
eigenvalue $\la_i$ one can try to use a recipe, discussed in
\cite{FaPe03} that might be called method of deformation of the
Hamiltonians, and goes as follows:

Consider the sum $p_1$ of all the eigenvalues of $N^*$, and the
Hamiltonian vector f\/ield $Y=-P_1 dp_1$. Then collect the
Hamiltonians f\/illing the Lenard--Magri recursion relations
\eqref{eq:1.1} in the Gel'fand--Zakharevich polynomial
\begin{gather}
\label{eq:1.7+} H(\la):=\la^n H_0+\la^{n-1}H_1+\cdots+\la
H_{n-1}+H_n
\end{gather}
and deform it repeatedly along the vector f\/ield $Y$, that is,
consider the polynomials
\begin{gather*}
  H'(\la)=\Lie{Y}H(\la),\qquad   H''(\la)=\Lie{Y}H'(\la),\quad \ldots  .
\end{gather*}
If, for some $n\ge 0$ it happens that the polynomial
$H^{(n+2)}(\la)$ identically vanishes in $\la$, (while
$H^{(n+1)}(\la)$ is {\em not} identically vanishing), then
evaluating the rational function
\begin{gather*}
  \dsl{\frac{H^{(n)}(\la)}{H^{(n+1)}(\la)}}\qquad \text{at}\qquad \la=\la_i
\end{gather*}
 provides us with a DN co-ordinate $g_i$ conjugate with $f_i\equiv\la_i$.

\subsection*{Remarks:}

1) When the eigenvalues $\la_i$ are no more functionally
independent, no general algorithms/recipes are known (at least to
the author of the present paper) to f\/ind \dncoo. However, when
the \Nij\ operator has a constant eigenvalue, say, $\la_n$, a case
that happens for the symmetric $SO(4)$ top herewith considered,
one can proceed as follows. One can at f\/irst apply the method
brief\/ly recalled above to construct a set of $(n-1)$ pairs of
\dncoo, associated with the $n-1$ co-ordinates
$\{\la_\al\}_{\al=1}^{n-1}$ def\/ined by the \Nij\ tensor. Then
one has to consider the distribution associated with the constant
eigenvalue $\la_n$, and try to integrate it (that is, to f\/ind
the functions $f_n$ and $g_n$ of equation~\eqref{eq:1.6}) by {\em
ad hoc} methods.

2) It should be stressed that, with respect to the pair $P_1, Q$,
the Hamiltonians do not f\/ill in a standard Lenard--Magri chain,
but rather a generalised one (see \cite{FMT98, FaPe03, Ma03} for
further details) of the form
\begin{gather*}
  Q dH_i=P_0 dH_{i+1}+p_i dH_1, \qquad i=1,\ldots, n,
\end{gather*}
where $p_i$ are (up to signs) the elementary symmetric polynomials
associated with the eigenvalues~$\la_i$. However, this invariance
relation is suf\/f\/icient to ensure separability of the
Hamilton--Jacobi equations.

3) Contrary to other methods for integrating of Hamilton
equations, and notably the method of Lax pairs, the bihamiltonian
setting herewith brief\/ly sketched provides, on general grounds,
somewhat poor information on the Jacobi separation relations --
that is, the relations tying pairs of separation co-ordinates with
the Hamiltonians $H_0,H_1,H_2,\ldots,H_n$ and the common Casimir
functions $\mathbf{C}=(C_1,\ldots, C_p)$.

However, the functional form of the separation relations can be
sometimes ascertained from bihamiltonian geometry. Indeed, if the
second Lie derivative of the GZ polynomial \eqref{eq:1.7+} with
respect to the transversal vector f\/ield $Z$ vanishes, then these
relations will be {\em
  affine} functions of the Hamiltonians and the Casimirs,
that is, they  will be given by expressions of the form
\begin{gather*}
  F_{1}^i(\la_i,\xi_i) H_1+\cdots
  +F_n^i(\la_i,\xi_i)H_n+G(\la_i,\xi_i;\mathbf{C})=0, \qquad i=1,\ldots, m.
\end{gather*}
Separation relations of this kind are often referred to as {\em
generalised} \St\ separation relations.

\section[The Euler-Manakov model]{The Euler--Manakov model}\label{sec.2}
In this section we will brief\/ly review the basic features of the
$SO(4)$ Euler--Manakov top.

The phase space is the (dual of) the Lie Algebra $\so(4)$,
identif\/ied\footnote{We are  actually identifying $\so(4)$ and
its dual.} with $4\times 4 $ antisymmetric matrices
\begin{gather*}
  M=\sum_{i<j=1}^4 m_{ij}(\Ee_{ij}-\Ee_{ji}),
\end{gather*}
where $\Ee_{ij}$ is the elementary matrix with $1$ at the
$(i,j)$-th place.

This six dimensional manifold  is naturally endowed with the Lie
Poisson structure, that, in the natural variables
$\mathbf{m}=\{\,m_{1,2},m_{1,3},m_{1,4},m_{2,3},m_{2,4},
m_{3,4}\,\}$ is represented by the matrix
\begin{gather*}
P_1=\left[ \begin {matrix} 0&-m_{2,3}&-m_{2,4}&m_{1,3}&m_{1,4}&0
\\
m_{2,3}&0&-m_{3,4}&-m_{1,2}&0&m_{1,4}
\\
m_{2,4}&m_{3,4}&0&0&-m_{1,2}&-m_{1,3}
\\
-m_{1,3}&m_{1,2}&0&0&-m_{3,4}&m_{2,4}
\\
-m_{1,4}&0&m_{1,2}&m_{3,4}&0&-m_{2,3}
\\
0&-m_{1,4}&m_{1,3}&-m_{2,4}&m_{2,3}&0
\end{matrix} \right].
\end{gather*}
The Hamiltonian is the quadratic function
\begin{gather}
\label{2.3} H_\CE=\frac12 \sum_{i<j=1}^4 {a_{ij}}m_{ij}^2,
\end{gather}
where the coef\/f\/icients $a_{ij}$ can be written as
\begin{gather*}
a_{ij}=J_l^2+J_k^2,\qquad\text{with}\qquad \{i,j,l,k\}\ \text{a
permutation of}\  \{1,2,3,4\}.
\end{gather*}
The Hamilton equations of motion (that is, the Euler equations for
the $SO(4)$ rigid body), are quadratic equations in the variables
$m_{ij}$ that depend parametrically on the coef\/f\/icients
$J_l^2$. For instance,
\begin{gather*}
\dsl{\frac{d}{dt}}{m_{12}}={J_1}^2(m_{1,3}m_{2,3}+m_{1,4}m_{2,4})
-{J_2}^2(m_{1,3}m_{2,3}+m_{1,4}m_{2,4})
\end{gather*}
and so on and so forth.

Complete Liouville integrability of the model is ensured following
well known facts.

1.~The rank of the $\so(4)$ Lie Poisson structure is $4$; its
Casimir functions are
\begin{gather*}
H_0=\sum_{i<j}m_{ij}^2,\qquad
C=m_{1,2}m_{3,4}+m_{1,4}m_{2,3}-m_{1,3}m_{2,4}.
\end{gather*}

2.~The second independent non-trivial constant of the motion for
$H_\CE$ is provided by the quadratic function
\begin{gather*}
K_\CE=\sum_{i<j=1}^4 b_{ij} m_{ij}^2,\quad
b_{ij}=J_l^2\,J_k^2,\qquad \text{with}\quad \{i,j,l,k\}\ \text{a
permutation of}\ \{1,2,3,4\}.
\end{gather*}

The Hamiltonian vector f\/ield $X$ associated with $H_\CE$ admits
a Lax representation with para\-meter \cite{BBEIM,Ma76}; indeed,  if
one considers the matrix $\Jj:={\rm diag}(J_1,J_2,J_3,J_4)$, and
forms the matrix
\begin{gather}
\label{2.6}
    L(\la)=\la\,\Jj^2+M,
\end{gather}
the Euler equations are equivalent to the Lax equations
\begin{gather*}
    \frac{d}{dt} L(\la)=[L(\la), B(\la)],
\end{gather*}
where $B(\la)=\Omega+\la\Jj$, and $\Omega$ is the `matrix of
angular velocities', i.e., def\/ined by $M$ via
$M=\Jj\,\Omega+\Omega\,\Jj$.

As it is well known, the integrals of the motion (as well as the
Casimirs of the Lie Poisson structure) are collected in the
characteristic polynomial of $L(\la)$. In particular, if one uses
the product $\rho\la$ as `eigenvalue parameter',  one gets
\begin{gather*}
\mathop{\rm Det}(L(\la)-\rho\la\mathbf{1})=
\la^4(P_4(\rho))+\la^2(\rho^2(H_0)+\rho\,H_1+H_2)+C^2,
\end{gather*}
where $P_4(\rho)=\prod\limits_{i=1}^4 (J_i^2-\rho)$,
$H_1=-2\,H_\CE$, $H_2=K_\CE$, and $H_0$, $C$ are the two Casimirs
of the Lie Poisson structure (the second one being the
Pfaf\/f\/ian of $M$.

As it was discovered in \cite{Bols89,Bols89a}, and independently
in \cite{MP96}, the Euler--Manakov equations of motion admit a
bihamiltonian formulation that can be described as follows. The
matrix $\Jj^2$ def\/ines a deformed commutator on the Lie algebra
$\so(4)$ as:
\begin{gather}
\label{2.9}
    [M_1,M_1]_{\Jj^2}:=[M_1\,{\Jj}^2 \,M_2,M_2\,{\Jj}^2\, M_1]=M_1\Jj^2 M_2-M_2\Jj^2M_1.
\end{gather}
The Lie Poisson structure associated with $[\cdot,\cdot]_{\Jj^2}$
provides a second Hamiltonian structure $P_2$ on~$\so(4)$.
Compatibility with the standard one is assured by the method of
augmented translations, i.e., by the fact that
$[\cdot,\cdot]_\la=[\cdot,\cdot]_{\Jj^2}-\la[\cdot,\cdot]$ is a
one-parameter family of commutators, that is, the Jacobi identity
holds identically in $\la$. The $6\times 6$ matrix representing
the second Poisson structure in the phase space variables
$\mathbf{m}=\{m_{ij}\}_{i<j=1,\ldots,4}$ can be easily found to be
\begin{gather*}
P_2=    \left[\begin{matrix}
0&-{J_1}^2m_{2,3}&-{J_1}^2m_{2,4}&{J_2}^2m_{1,3}&{J_2}^2m_{1,4}&0
\\
{J_1}^2m_{2,3}&0&-{J_1}^2m_{3,4}&-m_{1,2}{J_3}^2&0&m_{1,4}{J_3}^2
\\
{J_1}^2m_{2,4}&{J_1}^2m_{3,4}&0&0&-m_{1,2}{J_4}^2&-m_{1,3}{J_4}^2
\\
-{J_2}^2m_{1,3}&m_{1,2}{J_3}^2&0&0&-{J_2}^2m_{3,4}&m_{2,4}{J_3}^2
\\
-{J_2}^2m_{1,4}&0&m_{1,2}{J_4}^2&{J_2}^2m_{3,4}&0&-m_{2,3}{J_4}^2
\\
0&-m_{1,4}{J_3}^2&m_{1,3}{J_4}^2&-m_{2,4}{J_3}^2&m_{2,3}{J_4}^2&0
\end{matrix} \right].
\end{gather*}
In particular, the $\so(4)$ Euler system is a Hamitonian vector
f\/ield also w.r.t.\ the Poisson ope\-ra\-tor $P_2$, def\/ined by
the deformed commutator \eqref{2.9}, with `second' Hamiltonian
the function $-(1/2)\,H_0={-(1/2)\sum\limits_{i<j}m_{ij}^2}$.
Moreover, a direct computation ensures the following:
\begin{prop}
The characteristic polynomial \eqref{2.9} $\CP(\la,\rho)={\rm
Det}(L(\la)-(\rho\la)\mathbf{1})$ of the Lax matrix $L(\la)$ is a
Casimir of the Poisson pencil $P_\rho=P_2-\rho P_1$, that is,
\begin{gather*}
P_ 2\big(d\, \CP(\la,\rho)\big)=\rho\,P_
1\big(d\CP(\la,\rho)\big)\quad\text{identically in}\quad \rho, \
\la.
\end{gather*}
In other words, the Pfaffian $C$ is a common Casimir of the two
structures, while the three functions $H_0$, $H_1$, $H_2$ satisfy
the GZ recurrence relations
\begin{gather*}
P_1d (H_0)=0,\qquad P_2 d(H_0)=P_1 d(H_{1}),\qquad P_2
d(H_1)=P_1d(H_{2}),\qquad P_2 d(H_2)=0.
\end{gather*}
\end{prop}
For the sequel of the paper, the following well known
considerations will be useful.

{\sloppy The Lie algebra $\so(4)$ is isomorphic to the direct sum
$\so(3)\oplus\so(3)$; a linear change of variables that explicitly
realises this isomorphism is the following:
\begin{gather*}
x_1=\dfrac{1}{\sqrt{2}}(m_{1,2}-m_{3,4}),\qquad
y_1=\dfrac{1}{\sqrt{2}}(m_{1,3}+m_{2,4}),\qquad
z_1=\dfrac{1}{\sqrt{2}}(m_{1,4}-m_{2,3}),
\\
x_2=\dfrac{1}{\sqrt{2}}(m_{1,2}+m_{3,4}),\qquad
y_2=\dfrac{1}{\sqrt{2}}(m_{1,3}-m_{2,4}),\qquad
z_2=\dfrac{1}{\sqrt {2}} (m_{1,4}+m_{2,3}).
\end{gather*}
In particular, the variables $\{x_i,y_i,z_i\}_{i=1,2}$ satisfy,
with respect to the standard Lie Poisson structure $P_1$ the
$\so(3)$ commutation relations:
\begin{gather*}
\{x_i,y_i\}_{P_1}=\sqrt{2}z_i,\qquad
\{x_i,z_i\}_{P_1}=-\sqrt{2}y_i,\qquad
\{y_i,z_i\}_{P_1}=\sqrt{2}x_i,\qquad i=1,2,
\end{gather*}
while Poisson brackets involving co-ordinates from dif\/ferent
$\so(3)$ subalgebras vanish, e.g.\ $\{x_1,z_2\}_{P_1}=0$ and so on
and so forth.

}

Under this co-ordinate change, the Euler Hamiltonian \eqref{2.3}
acquires the form
\begin{gather}
\label{2.13} H_\CE= 2\mu_4x_1x_2+2\mu_3y_1y_2+2\mu_2z_1z_2
+\mu_1\left({y_1}^2+{y_2}^2+
{x_2}^2+{x_1}^2+{z_1}^2+{z_2}^2\right),
\end{gather}
where the new constants $\mu_i$ are related with the $J_i$'s as
follows:
\begin{gather*}
{J_1}^2=-\mu_4+\mu_1-\mu_3-\mu_2,\qquad
{J_2}^2=\mu_3-\mu_4+\mu_1+\mu_2,\nonumber
\\
{J_3}^2=\mu_1-\mu_3+\mu_4+\mu_2,\qquad
{J_4}^2=-\mu_2+\mu_1+\mu_3+\mu_4.
\end{gather*}

One can notice that the Hamiltonian \eqref{2.13} is the sum of a
multiple of the Casimir function~$H_0$ of the standard
Lie--Poisson structure, and the classical analogue
\begin{gather*}
H_{XYZ}=2(\mu_4x_2x_1+\mu_3y_1y_2+ \mu_2 z_1z_2)\end{gather*} of
the Hamiltonian of a ($2$-site) $XYZ$ Heisenberg model.

We will, in the remainder of the present paper, consider a special
case, namely the one that goes under the name of $XXZ$ model, that
is we will study the case $\mu_4=\mu_3$. From the point of view of
the Euler rigid body in $\so(4)$, this is tantamount to consider a
rigid body with two principal inertia moments ($J_2^2$ and
$J_3^2$) that are equal.

\section{The symmetric (or $\mathbf{\emph{XXZ}}$) Euler systems}\label{sec.3}
In the case $\mu_4=\mu_3$, the non-trivial part of the Hamiltonian
reads
\begin{gather*}
  H_{XXZ}=2\,\mu_3(x_2x_1+y_1y_2)+2\mu_2z_1\,z_2.
\end{gather*}
This fact suggests to choose linear co-ordinates in
$\alg=\so(3)\oplus\so(3)$ adapted to the symmetries of~$H_{XXZ}$;
the choice we will follow in the sequel will be to consider the
sixtuple $\{u_1,v_1,z_1,u_2,v_2,z_2\}$ related with the standard
$\so(3)\oplus\so(3)$ co-ordinates $\{x_k,y_k,z_k\}_{i=1,2}$ by
\begin{gather*}
u_k=x_k+i y_k,\qquad v_k=x_k-i y_k,\qquad k=1,2.
\end{gather*}
In these co-ordinates, the characteristic polynomial of the Lax
matrix $\hat{L}(\la)$ associated with the problem (that is, the
one given in \eqref{2.6}, with $\mu_4=\mu_3$) has the expression
\begin{gather*}
\la^4(\hat{P}_4(\rho))+\la^2(\rho^2
\CH_0+\rho\,\CH_1+\CH_2)+\frac14C_2^2,
\end{gather*}
where
\begin{gather}
\CH_0=u_1v_1+v_2u_2+{z_1}^2+{z_2}^2, \qquad
C_2=u_2v_2+{z_2}^2-u_1v_1-{z_1}^2,\nonumber
\\
\CH_1=-2\mu_3(u_2v_1+v_2u_1) -4\mu_2z_1z_2-2\mu_1\CH_0,\nonumber
\\
\CH_2= \mu_1^2\CH_0+4\mu_1\mu_2
z_1z_2+2\mu_3(\mu_1\!+\!\mu_2)(v_2u_1\!+\!u_2v_1)
+\mu_2^2({z_1}^2+{z_2}^2-v_1u_1-v_2u_2)\nonumber
\\ \phantom{\CH_2=}
{} -2 \mu_3^2(z_1-z_2t)^2. \label{eq:3.3}
\end{gather}
The explicit expressions of the Poisson tensors (in the new
co-ordinates) are, respectively,
\begin{gather}
\label{eq:3.4}
P_1=\left[\renewcommand\arraycolsep{4pt}\renewcommand\arraystretch{1.2}\begin{matrix}
A_1&\mathbf{0}
\\
\mathbf{0}&-A_2
\end{matrix}\right]\qquad\text{with}\qquad
A_i=\left[\renewcommand\arraycolsep{4pt}\renewcommand\arraystretch{1.2}\begin{matrix}
0&2z_i&-u_i
\\
-2z_i&0&v_i
\\
u_i&-v_i&0\end{matrix} \right]
\end{gather}
and $P_2=\mu_1\,P_1+\Delta, $ with $P_1$ still given by
\eqref{eq:3.4}, and
\begin{gather*}
\Delta=
\mu_2\left[\renewcommand\arraycolsep{4pt}\renewcommand\arraystretch{1.2}\begin{matrix}
0&2z_2&0&0&0&u_1
\\
&0&0&0&0&-v_1
\\
&&0&u_2&-v_2&0
\\
&&&0&-2z_1&0
\\
&&*&&0&0
\\
&&&&&0
\end{matrix}\right]+\mu_3\left[\renewcommand\arraycolsep{4pt}\renewcommand\arraystretch{1.2} \begin{matrix}
0&0&-u_2&0&2(z_2-z_1)&-u_2
\\
&0&v_2&2(z_1-z_2)&0&v_2
\\
&&0&-u_1&v_1&0
\\
&&&0&0&u_1
\\
&&*&&0&-v_1
\\
&&&&&0\end{matrix} \right].
\end{gather*}
(We indicate with a $*$ the lower diagonal part of these
antisymmetric tensors.)

The Hamiltonian vector f\/ield $X_1$ (i.e., up to a numeric
factor, the Euler--Manakov $SO(4)$ f\/ield in the rotationally
symmetric case), generated under $P_1$ by the Hamiltonian $\CH_1$
is explicitly given, in these new co-ordinates, by
\begin{gather}
X_1=4(\mu_2u_1z_2-\mu_3u_2z_1)\ddd{}{u_1}
-4(\mu_2v_1z_2-\mu_3v_2z_1)\ddd{}{v_1}
-4(\mu_2u_2z_1-\mu_3u_1z_2)\ddd{}{u_2}\nonumber
\\ \phantom{X_1=}
{}+4(\mu_2v_2z_1-\mu_3v_1z_2)\ddd{}{v_2} +2 \mu_3(u_2v_1-
u_1v_2)\biggl(\ddd{}{z_1}+\ddd{}{z_2}\biggr). \label{eq:3.6x}
\end{gather}
We will study the SoV problem for this Hamiltonian vector f\/ield,
within the scheme of SoV for GZ systems of \cite{FaPe03}, resumed
in Section \ref{sec.1}. Namely we have to:

1.~Find a suitable transversal vector f\/ield $Z$;

2.~Consider the Poisson operators $P_1, Q:=P_2-X_1\wedge Z$, as
well as their restrictions to the generic symplectic leaves;

3.~Find the DN co-ordinates associated with these restrictions;

4.~Find the Jacobi separation relations linking pairs of DN
co-ordinates and the Hamiltonians.

One can check that  a suitable transversal vector f\/ield for the
problem is given by
\begin{gather*}
  Z=\frac{1}{2u_1}\frac{\del}{\del v_1}+\frac{1}{2u_2}\frac{\del}{\del v_2}.
\end{gather*}
Namely,  $Z$ is a symmetry of $P_1$, that is, $\Lie{Z}(P_1)=0$.
Moreover,
\begin{gather*}
\Lie{Z}(\CH_0)=1, \qquad \Lie{Z}(C_2)=0,
\end{gather*}
and one indeed can check that the bivector
\begin{gather*}
  Q=P_2-X_1\wedge Z
\end{gather*}
turns out to be a (generically rank $4$) Poisson operator
compatible with $P_1$ that admits $\CH_1$ and~$C_2$ as Casimir
functions, that is, shares with the Poisson tensor $P_1$,
associated with the standard Lie algebra structure of
$\so(4)=\so(3)\oplus\so(3)$ the same symplectic leaves.

\medskip

\noindent {\bf Remark.} For further use, we notice that a direct
computation shows that the second Lie derivative of the
characteristic polynomial ${\rm
  Det}(L(\la)-(\rho\la)\mathbf{1})$ w.r.t.\ the transversal vector f\/ield~$Z$ vanishes as well,
\begin{gather}
\label{eq:3.x6} \Lie{Z}\left(\Lie{Z}({\rm
Det}(L(\la)-(\rho\la)\mathbf{1})\right)=0,
\end{gather}
that is, the condition for \St\ separability is fulf\/illed.

From the theoretical framework recalled in Section~\ref{sec.1} we
know that the symplectic leaves of~$P_1$ are four dimensional
manifolds endowed with a pair of compatible Poisson structure,
i.e., the restrictions of $P_1$ and of $Q$.

These four dimensional symplectic leaves $\CS$ are obtained
f\/ixing the values $\{\CH_0,{C_2}\}$ of these common Casimir
function. Furthermore, thanks to the explicit expressions given in
the f\/irst line of \eqref{eq:3.3}, in (open sets of) these
symplectic leaves one can use, as co-ordinates, the four
parameters
\begin{gather*}
{\uu}=\{u_1,z_1,u_2,z_2\},
\end{gather*}
since one can express the co-ordinates $v_1, v_2$ as follows:
\begin{gather*}
v_1=\frac12\,{\frac {\CH_0-C_2-2\,{z_1}^2}{u_1}},\qquad
v_2=\frac12\,{\frac {\CH_0+C_2-2z_2^2}{u_2}}.
\end{gather*}
The restrictions $\Pp$ and $\Qq$ of the Poisson structures $P_1$
and $Q$ to the leaf $\CS$ are represented by $4\times 4$ matrices
that have, in these co-ordinates, the explicit expressions
\begin{gather*}
\Pp=
\left[\renewcommand\arraycolsep{4pt}\renewcommand\arraystretch{1.2}\begin{matrix}
0&-u_1&0&0
\\
&0&0&0
\\
&*&0&u_2
\\
&&&0
\end{matrix} \right], \qquad
\Qq=\left[\renewcommand\arraycolsep{4pt}\renewcommand\arraystretch{1.2}\begin{matrix}
0&-(\mu_3u_2+\mu_1u_1)&0&\mu_2u_1-\mu_3u_2
\\
&0&\mu_2u_2-\mu_3u_1&0
\\
&*&0&\mu_1u_{{2}}+\mu_3u_1
\\&&&0
\end{matrix} \right].
\end{gather*}
The transpose Nijenhuis operator $N^*=\Pp^{-1}\Qq$ is given by
\begin{gather}
\label{ultima} \Nn^*=
\left[\renewcommand\arraycolsep{4pt}\renewcommand\arraystretch{1.2}\dsl{\begin{matrix}
\mu_3\dsl{{\frac {u_2}{u_1}}}+\mu_1&0&+\mu_2\dsl{{\frac
{u_2}{u_1}}-\mu_3}&0
\\
0&\mu_3\dsl{{\frac {u_2}{u_1}}}+\mu_1&0&\mu_3\dsl{{\frac
{u_2}{u_1}}}-\mu_2
\\
{\mu_2\dsl{\frac {u_1}{u_2}}}-\mu_3&0&\mu_3\dsl{{\frac
{u_1}{u_2}}}+\mu_1&0
\\
0&\mu_3\dsl{{\frac {u_1}{u_2}}}-\mu_2&0&\mu_3\dsl{{\frac
{u_1}{u_2}}}+\mu_1
\end{matrix}}\right].
\end{gather}
Its eigenvalues are
\begin{gather*}
\la_1=\mu_1+\mu_2,\qquad
\la_2=\mu_1-\mu_2+\mu_3\left(\dsl{\frac{u_1}{u_2}+\frac{u_2}{u_1}}\right).
\end{gather*}

From the general theory, we know that $\la_2$ is one of the DN
co-ordinates we are looking for (which we need to complement with
its conjugate co-ordinate $\xi_2$), while we need to f\/ind both
canonical co-ordinates relative to the constant eigenvalue
$\la_1$.

The problem of f\/inding the canonical co-ordinate conjugated to
the non--constant eigenvalue $\la_2$ can be dealt with the idea of
deforming the Hamiltonian polynomial. Indeed we consider the sum
$p_1$ of the eigevalues of $\Nn$,
$p_1=2\mu_1+\mu_3(\dsl{{u_1}/{u_2}+{u_2}/{u_1}})$, and  the vector
f\/ield $Y=-\Pp d p_1$; it is given by
\begin{gather*}
  Y=\mu_3 G(\uu)\left(\ddd{}{z_1}+\ddd{}{z_2}\right),
\end{gather*}
where the function $G(\uu)$ is given by
\begin{gather}
\label{eq:3.10x} G(\uu)={\frac {u_2}{u_1}}-{\frac {u_1}{u_2}},
\end{gather}
and is connected with the eigenvalue $\la_2$ by
\begin{gather}
  \label{eq:3.9x}
  G(\uu)^2=4\,+\left(\dsl{\frac{\la_2-\mu_1+\mu_2}{\mu_3}}\right)^2.
\end{gather}
Now we iteratively apply the vector f\/ield $Y$ to the polynomial
containing the relevant Hamiltonians, that is
\begin{gather*}
\CH(\rho)=\rho^2\,\CH_0+\rho\,\CH_1+\CH_2.
\end{gather*}
By means of direct computation one can check that $\Lie{Y}(\CH)$
factors as
\begin{gather*}
\Lie{Y}(\CH)=\frac{4 \mu_3
(\rho-\mu_1-\mu_2)}{u_1\,u_2}\,G(\uu)\,L(\uu)
\end{gather*}
with
\begin{gather*}
L(\uu)= \mu_3 \left(z_2{u_1}^2+z_1{u_2}^2 \right) -
 \mu_2u_1u_2\left(z_1+z_2 \right).
\end{gather*}
Since, quite obviously $\Lie{Y}(u_1\,u_2)=0$, and, thanks to
\eqref{eq:3.9x},
\begin{gather*}
\Lie{Y}(G(\uu))={\parpo{G(\uu)}{\la_2}}_\Pp=0,
\end{gather*}
we are left, for the computation of the second Lie derivative of
$\CH$, with the computation of $\Lie{Y}(L(\uu))$. It gives
\begin{gather*}
\Lie{Y}(L(\uu))=\mu_3G(\uu)\big(u_1\,u_2\, F(\uu)\big),
\end{gather*}
where
\begin{gather*}
F(\uu)=-2\mu_2+{\frac
{({u_1}^2+{u_2}^2)\mu_3}{u_1u_2}}=\la_2-(\mu_1+\mu_2).
\end{gather*}
So, the third Lie derivative of $\CH$ w.r.t. $Y$ vanishes, and so
the function
\begin{gather*}
  \xi_2=\frac{\Lie{Y}({\CH})}{\Lie{Y}(\Lie{Y}(\CH))}\bigg\vert_{\rho=\la_2}=
-\frac1{\mu_3\,u_1\,u_2}\biggl(\frac{L(\uu)}{G(\uu)\,F(\uu)}\biggr)
\end{gather*}
is the  DN co-ordinate conjugated to $\la_2$ we were looking for.

The two functions $F(\uu)$ and $G(\uu)$ will play a role in the
last task we will deal with, that is, the determination of the
Jacobi separation relations.

Next we turn to consider the problem of f\/inding DN co-ordinates
associated with the constant eigenvalue $\la_1=\mu_1+\mu_2$  of
the tensor $N^*$ of equation~\eqref{ultima}. Being this a constant
eigenvalue, we have to f\/ind `by hands' the associated DN
co-ordinates. Fortunately enough the expression of the operator
$N^*-\la_1$ is given by
\begin{gather*}
\left[\renewcommand\arraycolsep{4pt}\renewcommand\arraystretch{1.2}
\begin{matrix} -\mu_2+\dsl{\frac {\mu_3u_2}{u_1}}&0&\dsl{\frac
{\mu_2u_2}{u_1}}-\mu_3&0
\\
0&-\mu_2+\dsl{\frac {\mu_3u_2}{u_1}}&0&-\mu_2+\dsl{\frac
{\mu_3u_2}{u_1}}
\\
\dsl{\frac {\mu_2u_1}{u_2}}-\mu_3&0&\dsl{\frac
{\mu_3u_1}{u_2}}-\mu_2&0
\\
0&\dsl{\frac {\mu_3u_1}{u_2}}-\mu_2&0&\dsl{\frac
{\mu_3u_1}{u_2}}-\mu_2
\end{matrix} \right].
\end{gather*}
Its kernel can be easily found to be generated by the two 1-forms
\begin{gather*}
\al_1=dz_2-dz_1,\qquad
\al_2=(\mu_3u_1-\mu_2u_2)du_1+(\mu_3u_2-\mu_2u_1 )du_2
\end{gather*}
that integrate, respectively,  to the functions
\begin{gather*}
\zeta_1=z_2-z_1,\qquad \theta_1=\frac12\mu_3\,{u_1}^2
-\mu_2\,u_1u_2+\frac12\mu_3\,{u_2}^2.
\end{gather*}
A direct computation shows that $
\{\zeta_1,\theta_1\}_{\Pp}=-2\theta_1, $ and so the DN co-ordinate
conjugated to~$\zeta_1$ is $\xi_1=-\dsl{(1/2)}\log\theta_1$. Thus,
we have found, on the generic common symplectic leaf $\CS$ of
$P_1$ and~$Q$, the desired set of DN co-ordinates
$(\zeta_1,\xi_1=-(1/2)\log{\theta_1}, \la_2,\xi_2)$. It can be
noticed that, along with the two Casimirs $C_1$, $C_2$, they
provide a set of co-ordinates in a Zariski open subset of the
phase space $\so(4)\simeq\so(3)\oplus\so(3)$.

What we are left with the determination of the Jacobi separation
relations, namely we have to seek for two relations of the form
\begin{gather}\label{3.xx}
\Phi_1(\mathbf{H};\zeta_1,\xi_1)=0,\qquad
\Phi_2(\mathbf{H};\la_2,\xi_2)=0,
\end{gather}
linking pairs of DN co-ordinates and the conserved quantities
$\mathbf{H}=\{C_2,\CH_0,\CH_1,\CH_2\}$ of \eqref{eq:3.3}.

Owing to the dif\/ferent ways the separation co-ordinates were
found, and, especially, the fact that one of the separating
momenta is an additional constant of the motion, we expect
that~$\Phi_1$ and~$\Phi_2$ have dif\/ferent functional dependence
on their variables; so, instead of trying to use the spectral
curve relations we directly seek for the Jacobi relations
\eqref{3.xx}, by means of explicit calculations. Since the
characteristic condition for \St\ separability is verif\/ied in
our case (see equation~\eqref{eq:3.x6}), we can look for Jacobi
relations that are af\/f\/ine functions in the Casimirs and the
Hamiltonians.

Let us f\/irst consider $\Phi_1(\mathbf{H};\zeta_1,\xi_1)=0$; in
this respect, one can notice that $\zeta_1$ is an additional
constant of the motion. Indeed, from the form of the Euler vector
f\/ield \eqref{eq:3.6x}, we easily ascertain that
$\{\CH_1,\zeta_1\}_{P_1}=0$ and a direct computation (or a careful
examination of the generalised Lenard relations associated with
$\Pp,\Qq$) shows that $\zeta_1$ commutes with $\CH_2$ as well.

So, there must be a functional relation between $\CH_0$, $C_2$,
$\CH_1$, $\CH_2$ and $\zeta_1=z_1-z_2$ alone, that is, we expect
$\Phi_1$ to be independent of $\xi_1$. Taking into account that
the elements $\mathbf{H}$ are quadratic functions of $z_1$, $z_2$,
we look for a relation of the form
\begin{gather*}
\Phi_1=\al \zeta_1^2+\CH_1+\be \CH_2+ \ga_1\CH_0+\ga_2C_2,
\end{gather*}
for some unknown {\em constants} $\al$, $\be_i$, $\ga_i$, $i=1,2$.
Indeed such a relation can be found with, respectively,
\begin{gather*}
\al=2{\frac {{\mu_3}^2-{\mu_2}^2}{\mu_1+\mu_2}}, \qquad
\be=\frac1{\mu_1+\mu_2}, \qquad \ga_1=\mu_1+\mu_2,\qquad \ga_2=0.
\end{gather*}

To f\/ind the second separation relation is slightly more
involved; still the idea is to look for a relation quadratic in
$\xi_2$, and af\/f\/ine in $\CH_0$, $C_2$, $\CH_1$, $\CH_2$, with
coef\/f\/icients that may depend non trivially on the co-ordinate
$\la_2$, i.e.\ a relation of the form
\begin{gather}
  \label{eq:3.18}
  \Phi_2=p(\la_2)\xi_2^2+q_1(\la_2)\CH_1+q_2(\la_2) \CH_2-\Psi(\la_2,\CH_0,C_2),
\end{gather}
for some functions $p$, $q_1$, $q_2$ that depend only on $\la_2$,
and for an unknown function $\Psi(\la_2;\CH_0,C_2)$, af\/f\/ine in
the $C_i$'s. After a direct computation one sees that the problem
can be solved, and that the second separation relation has the
form \eqref{eq:3.18}, with
\begin{gather}
  \label{eq:3.19}
  q_1=\la_2,\qquad q_2=1,\qquad p=-2\mu_3^2(F(\uu)^2
G(\uu)^2),\qquad \Psi=\la_2^2 \CH_0-\mu_3
  F(\uu) G(\uu)\,
C_2,
\end{gather}
where the functions $G(\uu)$ and $F(\uu)$ are given respectively
by \eqref{eq:3.10x} and \eqref{eq:3.9x}.

It can be noticed that the relation \eqref{eq:3.19} is quadratic
in the momentum $\xi_2$, and is algebraic~-- rather that
polynomial~-- in the co-ordinate $\la_2$, owing to the relation
\eqref{eq:3.9x}.

The question whether these techniques might be useful in the study
of the general $\so(4)$ Euler--Manakov top, that is, the case
$\mu_3\neq \mu_4$, is still under investigation. What is still
lacking, in this general case, is the determination of the
suitable deformation vector f\/ield $Z$; it is conceivable that a
careful use of the theory of elliptic function may provide the
answer.

Also, the connection of the results herewith presented with the
Clebsh model, according to an isomorphism described in \cite{Bo86}
might be worth of further investigations\footnote{I thank one of
the referees for addressing this point.}, as well as the link with
the setting, within the bihamiltonian theory, of the Separation of
Variables problem for Lagrange tops presented in \cite{MoTo02}.

\subsection*{Acknowledgments} This work was partially supported by the
European Community through the FP6 Marie Curie RTN {\em ENIGMA}
(Contract number MRTN-CT-2004-5652), and by the European Science
Foundation  project {\em MISGAM}. Thanks are due to the anonymous
referees for their useful remarks.

\pdfbookmark[1]{References}{ref}
\LastPageEnding
\end{document}